# Protection of New York City's Urban Fabric With Low-Cost Textile Storm Surge Barriers

(Short version for Arxiv)

**Alexander A. Bolonkin**
C & R, 1310 Avenue R, Suite 6-F,
Brooklyn, New York 11229, USA
e-mail: abolonkin@gmail.com

**Richard B. Cathcart**
Geographos
1300 West Olive Avenue, Suite M
Burbank, California 91506, USA

**Abstract**

Textile storm surge barriers, sited at multiple locations, are literally extensions of the city's world famous urban fabric—another manifestation of the dominance of the City over local Nature. Textile Storm Surge Barriers (TSSB) are intended to preserve the City from North Atlantic Ocean hurricanes that cause sea waves impacting the densely populated and high-value real estate, instigating catastrophic, and possibly long-term, infrastructure and monetary losses. Complicating TSSB installation macroproject planning is the presence of the Hudson and other rivers, several small tidal straits, future climate change and other factors. We conclude that TSSB installations made of homogeneous construction materials are worthwhile investigating because they may be less expensive to build, and more easily replaced following any failure, than concrete and steel storm surge barriers, which are also made of homogeneous materials. We suppose the best macroproject outcome will develop in the perfect Macro-engineering planning way and at the optimum time-of-need during the very early 21$^{st}$ Century by, among other groups, the Port Authority of New York and New Jersey. TSSB technology is a practical advance over wartime harbor anti-submarine/anti-torpedo steel nets and rocky Churchill Barriers.

**Introduction**

For our purposes, a storm surge is an above normal rise in sea level accompanying a hurricane passing near or over New York City. New York City has endured North Atlanti1c Ocean storm surges caused by hurricanes—for example, lower Manhattan Island was flooded on 3 September 1821 by a Category 3 storm that made landfall near today's JFK Airport. The earliest calendar date of a hurricane affecting the State of New York's coast is 4 June, while the latest is 13 November. "September is the month of greatest frequency for storms of tropical origin, although the storms of greatest intensity tend to arrive later in the hurricane season…".[1] North Atlantic Ocean hurricane activity will probably not be significantly different from that of the 20$^{th}$ Century since all future Atlantic Ocean tropical storm activity will critically stem directly from the warming of the tropical Atlantic Ocean relative to that of the combined Indian Ocean-Pacific region.[2]

Indeed, recent climate studies suggest that "…North Atlantic hurricane activity is greater during La Nina years and suppressed during El Nino years…due primarily to increased vertical wind shear in strong El Nino years hindering hurricane development".[3] (An excellent summary of current scientific knowledge about tropical storms is available in Patrick J. Fitzpatrick's HURRICANES: A REFERENCE HANDBOOK (2nd Edition, 2005, 412 pages.)

Some previous cases of weak storm surge inundations in New York are shown below (Figs. 4 -10A – 10D).

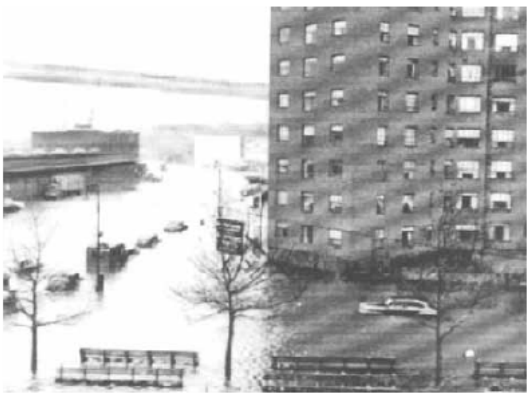
FIGURE 4-10A  Lower East Side, Manhattan. November 24, 1950.

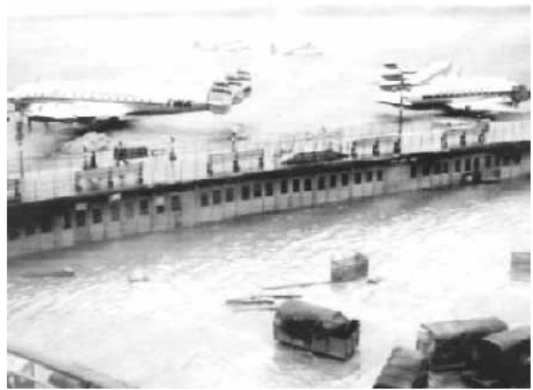
FIGURE 4-10B  La Guardia Airport, Queens. November 25, 1950.

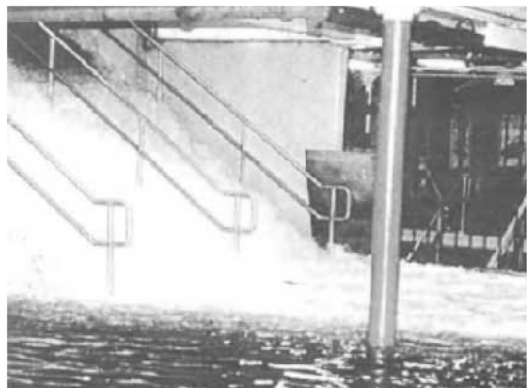
FIGURE 4-10C  Hoboken PATH Station, New Jersey. 1992 Nor'easter.

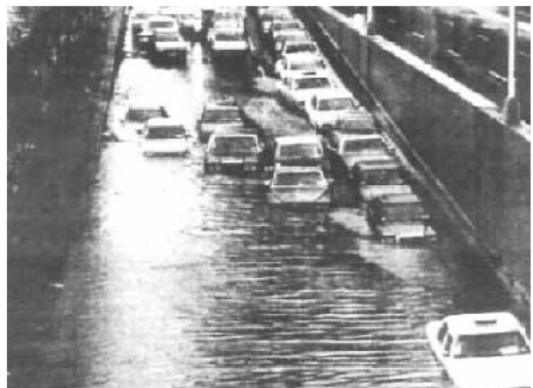
FIGURE 4-10D  FDR Drive Northbound at 80th Street, East River, Manhattan. 1992 Nor'easter.

Source: New York City Office of Emergency Management

North Atlantic Ocean hurricanes are rated according to wind speed by the Saffir-Simpson Scale first offered in 1969 by a civil engineer born in New York City, Herbert Saffir, and by a meteorologist, Robert Simpson.  Weather forecasters and infrastructure emergency managers in the USA utilize the Saffir-Simpson Scale to anticipate macro-problems for urban fabrics, which even extends into the countryside for the purpose of estimating the potential damage and flooding that might be stimulated by a hurricane's landfall.[4] Managers and other decision makers desire to know with some certainty whether either an evacuation or a shelter-in-place public announcement is required and what boroughs of the New York City are likely to be most impacted as well as for the storm impact period of duration—in other words, New York City's exposure and its weather sensitivity economically. Hurricanes normally weaken after landfall and the weakening is sometimes evidenced by potentially destructive short-term phenomena (tornadoes)! Coastal hills and cities skylines—Manhattan Island, covered as it is with tall buildings and other structures, including landfills[5], is almost the equivalent of a geological formation—can influence the dynamics of such storms through blocking and resultant disruption of the storm wind circulation, causing torrential rain in the urbanized and rural coastal zone.[6] The near-surface winds are decelerated by increased surface friction with natural and made landforms as well as artificial structures.[7] Construction mistakes of a landmark Manhattan skyscraper, the 278 m-high Citigroup Center completed in 1977, necessitated costly special reinforcement after it occupancy to successfully resist impacting hurricane winds.

New York City's meteorological vulnerability is a subject still being vigorously debated by all stakeholders. It is possible that abrupt global climate change could cause a noticeable rise in local sea levels, thereby aggravating the situation as regards hurricane impacts. All existing and planned infrastructure embodies the extant Macro-engineering design criteria codifying today's technology and science. Even so, there are few successful macroprojects. Any macroproject to protect New York City from hurricane storm surges must be indivisible: "Thus, for a [macroproject]…to proceed it must be in the best interests of all key participants, *i.e.*, it must be and *remain* superior to any other practical courses open to any of these key parties, including doing nothing."[8] All protective macroprojects must have the capability to endure not only hurricanes but earthquakes[9] and terrorist attacks such as that which occurred on 11 September 2001. A focus of planning in the metropolitan region since its inception in 1921, the Port Authority of New York and New Jersey which oversees seaport operations and associated operations must play a leading role in the planning and construction of the several required hurricane storm surge barriers needed to shield New York City and adjacent urban fabrics.[10] We assume that impermeable barriers like the Canso Causeway across the Strait of Canso, Nova Scotia, linking Cape Breton Island with Canada's mainland or the Churchill Barriers in the Orkney isles built to protect the UK's Navy's Scapa Flow anchorage.

**NYC Storm Surge Barrier Purpose**

Storm surges, the flooding induced by surface wind stresses and the barometric pressure reduction associated directly with hurricanes and northeasters, are a threat to New York City's infrastructure as well as other famous cities. The Thames Barrier in the UK and the Delta Project in The Netherlands were designed with heights to exceed the surge elevations of certain design storms. However, these barriers will assuredly become more ineffective as global and regional sea level rises during the 21$^{st}$ Century. Computer modeling of storm surge flooding of any urban fabric is rife with uncertainties.[11] Our TSSB (Textile Storm Surge Barrier) macroproject is meant to exclude only storm surges predicted to affect New York City's infrastructure.[12] Unlike the Thames Barrier and the Delta Project, the TSSB can be easily elevated (through refurbishment) to accommodate future regional sea level rise.[13] Here, we examine the potential effectiveness of hypothetical TSSB installations emplaced at The Narrows, the upper East River near

College Point (near the Bronx-Whitestone Bridge opened to vehicle traffic in 1939) and at the mouth of Arthur Kill near Perth Amboy, New Jersey; in other words, we hope to prevent damaging hurricane and northeaster storm surges from entering the New York and New Jersey metropolitan region by way of Lower New York Bay, Raritan Bay and Long Island Sound. Much of the land facing these bodies—estimated at ~260 km$^2$—of seawater are <3 m ASL. The TSSB sites we have selected are identical to those suggested by the Stony Brook Storm Surge Research Group currently led by Professor Malcolm J. Bowman.[14] (GOTO: http://stormy.msrc.sunysb.edu/sbss-Banner.htm .) Emplacement of three TSSB facilities, which would remain open to shipping until closed prior to expected hurricane and northeaster storm surge events, means that, like other concretized blockages, the TSSB will retain ~133 km$^2$ of water surface.

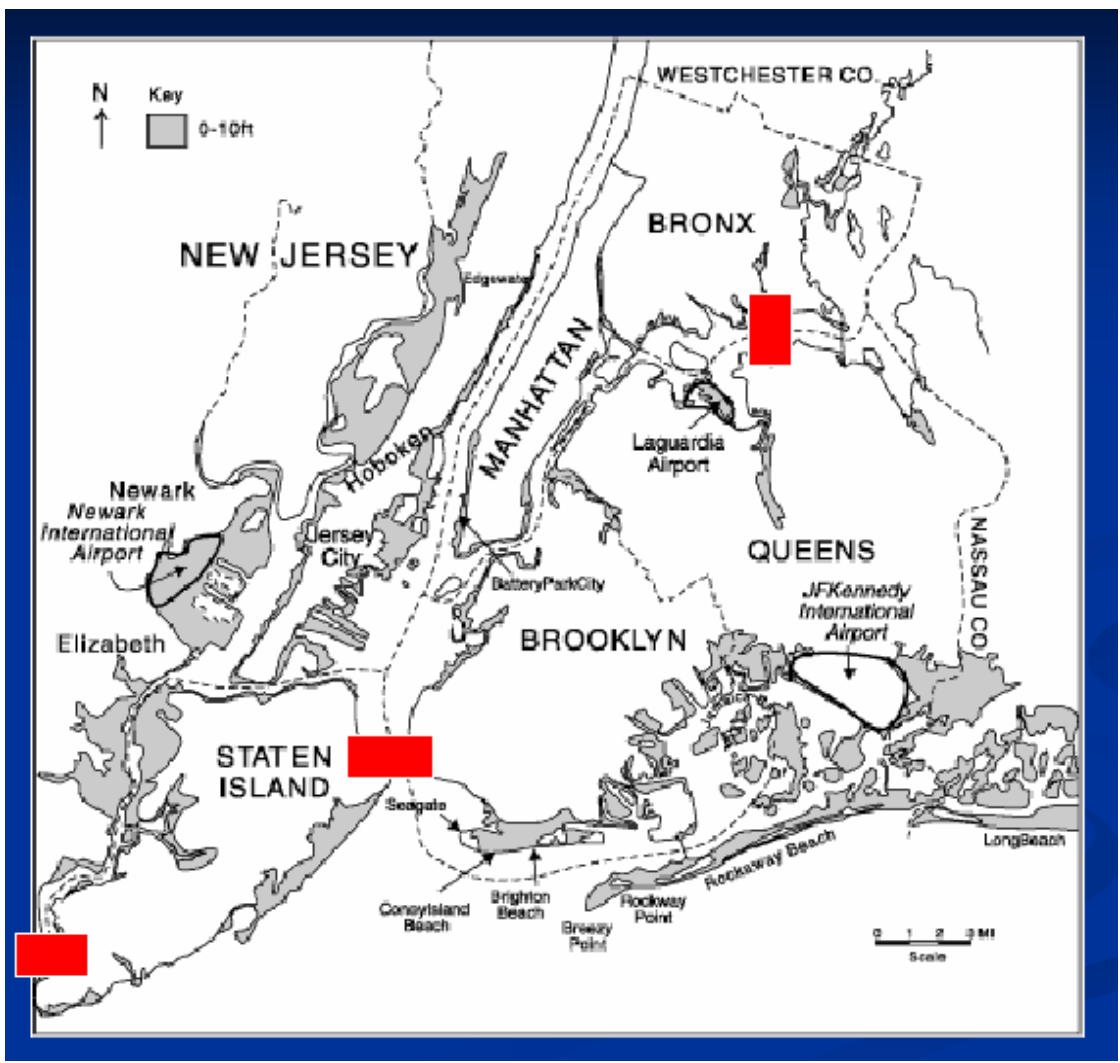

**Fig. 5**. The Dams are offered by M. Bowman.

During the 1971-72 of investigations preparatory to a final builder-owner decision on the Thames Barrier's design, an ingenious fabric dam was considered after its submission by Andrew Noel Schofield.[15] At page 55 of Stuart Gilbert and Ray Horner's THE THAMES BARRIER (1984, Thomas Telford Ltd) Schofield's dam is described: "A vast fabric sheet was to rest on the bed of the river (it could be removed for maintenance during the summer non-surge season). The upriver edge of the sheet would be linked through running gear to a heavy hawser or cable, both sheet and cable spanning the river. The sheet could be raised by winches on the river banks hauling in the cable. Other cables attached to strong anchors in the bed of the river downriver would be secured to the main cable across the river. By underdraining below the sheet and pumping, the lower edge of the sheet would be gripped between the water pressure and the river bed. As the surge came in the depth of water and holding pressure would increase. The underdrainage would be formed by perforated plastic pipes inserted in layers of clean gravel dropped into a trench dredged across the river bed. Any silt settling on the sheet could be tipped off it by raising the sheet on the outgoing tide. This was an unusual solution and had its attractions…". The environmental impact of the use of fabric and textiles is a little-examined field; the short and long-term environmental impact of textile production has been studied.[16] There is only one published report available that deals with something quite similar environmentally to Schofield's 30+ year-old macroproject proposal.[17] [Here, it is necessary to note that fiber-reinforced polymer (FRP) employment is increasing worldwide. For example, the 114 m-long cable-stayed Aberfeldy Footbridge over the Tay River in Scotland, finished in 1990, has a pedestrian deck of FRP composite panels and a supporting pylon composed of FRP and the cable stays transmitting the deck's dead and live loads to the pylon's top are comprised of woven Kevlar.]

At certain work sites, macro engineers will have to span nearly 1.6 km of tidal strait water! (The Thames Barrier protecting London is only 523 m-long.) At the Narrows, it may be possible to form a TSSB "add-on" to the 1964 Verrazano-Narrows Bridge. The so-called East River, which is actually a tidal strait, is the site of the USA's first major tidal-power macroproject; ultimately, ~100 underwater turbines, each of ~35 kwe generating capacity, will be installed in the East River. Such a facility will be affected adversely by closure of the planned TSSB! No seawater flow, no electricity output! Closure of the TSSB will not damage the facility whatsoever. However, what is the point of worrying about a temporary power shutdown when the urban fabric is at risk of flooding and extreme, long-lasting physical damage? Kate Ascher's ANATOMY OF A CITY (2005, The Penguin Press, 228 pages) well illustrates that complex urban fabric, but without any cautionary mention of hurricane storm surge threats (or suggested Macro-engineering cures)! Reference #7 cited below shows vividly the progress that has been made in modeling cities—see, for example, old-fashion physical models of New York City as it looked by 1992.[18]

**Physics of Maximum Width TSSB (The Narrows)**

The Narrows separates the New City boroughs of Staten Island and Brooklyn. It is the main channel through which the post-Ice Age Hudson River drains.[19] "The Hudson estuary has an extremely low stability ratio (0.4) which is defined as the ratio of the volume of an estuary (in $km^3$) to the mean flow of freshwater in $m^3/s$ into the system….

This means that it has a very rapid response to storms and hurricanes, with the salt front moving long distances very rapidly…".[20] Episodic hurricanes deliver sediment to the Hudson River estuary since the estuary's geologic creation less than 13,000 years ago.[21] The possible worksite for our TSSB beneath or very near the Verrazano-Narrows Bridge offers a Quaternary alluvium geology typified by the "Harbor Hill Terminal Moraine" formation. (The suspension bridge spanning 2040 m of seawater was built during 1959-64, and cost then approximately $60 millions.) We presume at TSSB located thereabouts wont exceed that cost incurred over forty years ago! And, we certainly reject an alleged estimate of the cost for total metropolitan anti-storm surge protection as "…tens of billions of dollars".[22] As we mentioned previously, there must be three macroprojects to effectively guard the metropolitan region from infrequent storm surge events because there is the potential for an infrastructure loss of utility amounting to an estimated $300 billions! The Stony Brook Storm Surge Research Group's *circa* 2002 recommendation for three barriers is controversial: "…its economic, cost/benefit and political feasibility and long-term environmental impacts are as yet entirely unknown. It would be a capital-intensive 'structural' solution…".[23]

Following is a brief presentation of the physics of the Textile Storm Surge Barrier as exemplified by a single installation at the Narrows.

## Description of innovation [25]-[30]

Current coast-protection dams are built from solid material (heaped stones, concrete, piled soil). They are expensive to emplace and, sometimes, are unsightly. Such dams require detailed on-site research of the surface and sub-surface environment, costly construction and high-quality building efforts over a long period of time (years). Naturally, the coast city inhabitants lose the beautiful sea view and ship passengers are unable to admire the city panorama of the partly hidden city (New Orleans, which is below sea level). The sea coastal usually has a complex geomorphic configuration that greatly increases the length and cost of dam protections.

Authors offer to protect seaport cities against hurricane storm surge waves, tsunamis, and other weather related-coastal and river inundations by new special design of the water and land textile dams.

The offered dam is shown in Fig. 6a below. One contains the floats 4, textile (thin film) 3 and support cables 5. The textile (film) is connected a top edge to the floats, the lower edge to a sea bottom. In calm weather the floats are located on the sea surface (Fig. 6a) or at the sea bottom (Fig. 6b). In stormy weather, hurricane, predicted tsunami the floats automatically raise to top of wave and defend the city from any rapid increase of seawater level (Fig. 6c).

This textile-based dam's cost-to-build is thousands of times cheaper then a massive concrete dam, and a textile infrastructure may be assembled in few months instead of years! They may be installed on ground surface around vital or important infrastructure objects (entries to subway tunnels, electricity power plants, civic airport, and so on) or around a high-value part of the city (example, Manhattan Island) if inundation poses a threat to the city (Fig. 7). These textile protections are mobile and can be relocated and installed in few days if hurricane is predictably moving to given city. They can defend the noted object or city from stormy weather inundation, tsunamis, and large waves of height up 10 and more m.

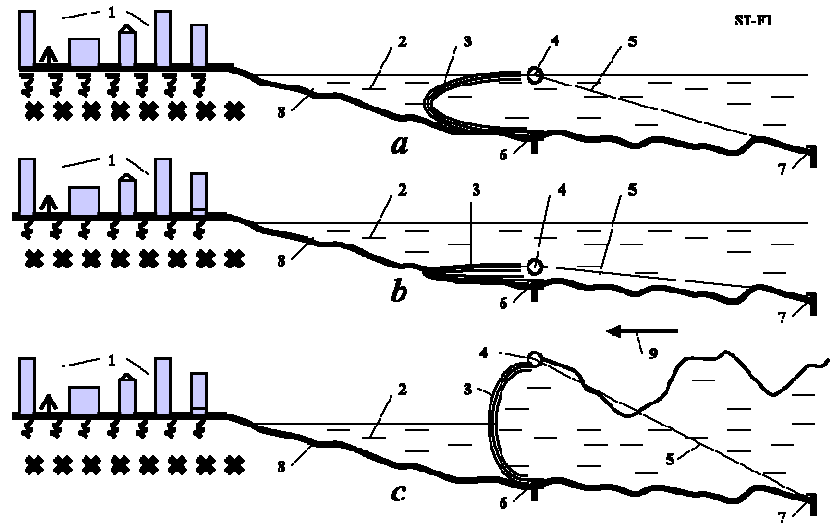

**Fig. 6.** Protection city against hurricane storm surge waves, tsunamis, and other weather related-coastal inundations by textile (film) membrane located in sea (ocean). (*a*) - position of membrane on a sea surface in a calm weather; (*b*) - position of membrane on a sea bottom in a calm weather; (*c*) - position of membrane in hurricane storm surge waves, tsunamis, and other weather related-coastal inundations. Notations: 1 - city, 2 - sea (ocean), 3 - membrane, 4 - float, 5 - support cable, 6 - connection of membrane to a sea bottom, 7 - connection of support cable to a sea bottom, 8 - sea bottom, 9 - wind.

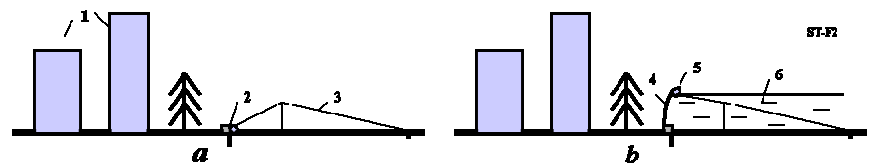

**Fig.7.** Protection city against hurricane storm surge waves, tsunamis, and other weather related-coastal inundations by textile (film) membrane located on ground surface. (*a*) - position of membrane on a ground surface in a compact form in a calm weather; (*b*) - position of membrane in hurricane storm surge waves, tsunamis, and other weather related-coastal inundations. Notations: 1 - city, 2 - membrane in the compact form, 3 - support cable, 4 - membrane, 5 - float, 6 -water, 7 - connection of support cable to a sea bottom.

The offered textile dam may be also used as a big source of electricity. They can be built as the dams in rivers and it is used as water dams for the electric station (Fig. 8).

They also can be used as the dams for an ebb - flow sea electric station (Fig. 9).

The membranes must be made from artificial fiber or a film. The many current artificial fibers are cheap, have very high safety tensile stress (some times more the steel!) and chemical stability. They can work as dam some tens years. They are easy for repair.

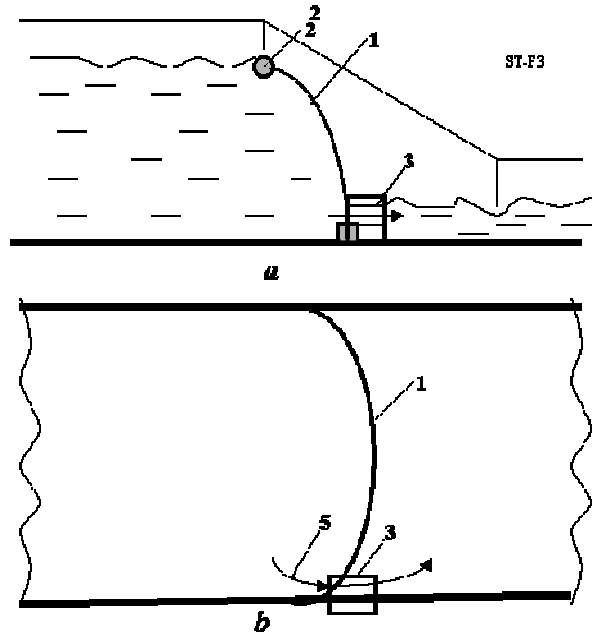

**Fig. 8.** Textile dam and electric station in a river. (*a*) side view; (*b*) - top view. Notations: 1 - textile dam, 2 - float, 3 - electric station, 4 - water flow.

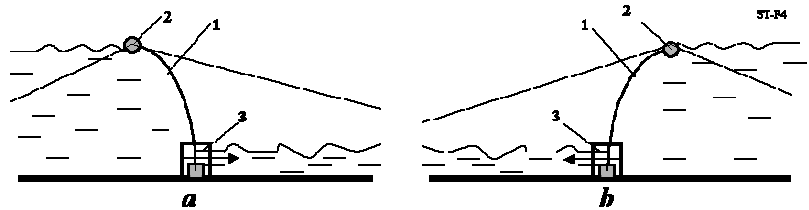

**Fig.9**. Tidal (ebb and flow) electric station with textile dam. (*a*) - ebb; (*b*) - flow. Notations: 1 - textile membrane, 2 - float, 3 - electric station.

**Theory and Computation**

1. **Force** $P$ [N/m$^2$] for 1 m$^2$ of dam is

$$P = g\gamma h, \qquad (1)$$

where $g$ = 9.81 m/s2 is the Earth gravity; $\gamma$ is water density, $\gamma$ =1000 kg/m3; $h$ is difference between top and lower levels of water surfaces, m (see computation in Fig. 10).

2. **Water power** $N$ [W] is

$$N = \eta g m h, \quad m = \gamma v S, \quad v = \sqrt{2gh}, \quad N = \eta g \gamma h S \sqrt{2gh}, \quad N/S \approx 43.453\eta h^{1.5}, \ [\text{kW/m}^2] \qquad (2)$$

where $m$ is mass flow across 1 m width kg/m; $v$ is water speed, m/s; $S$ is turbine area, m$^2$; $\eta$ is coefficient efficiency of the water turbine, $N/S$ is specific power of water turbine, kW/m$^2$.

Computation is presented in Fig. 11.

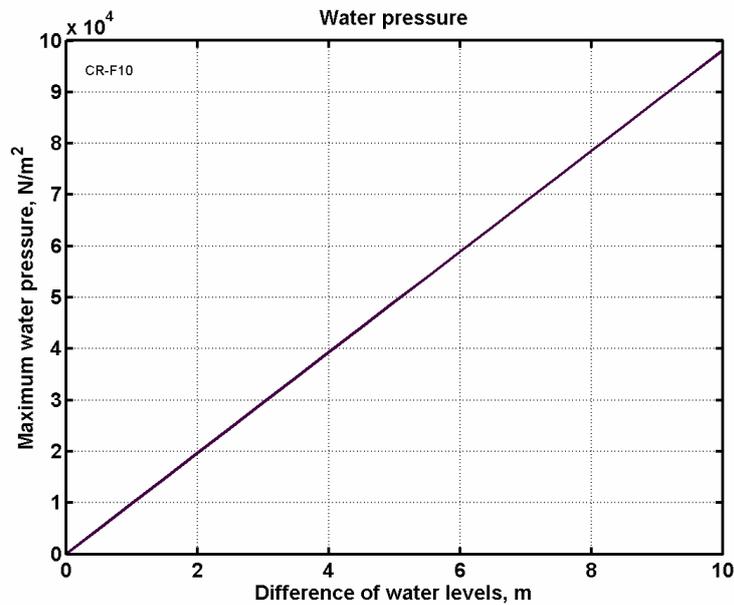

**Fig. 10**. Water pressure via difference of water levels

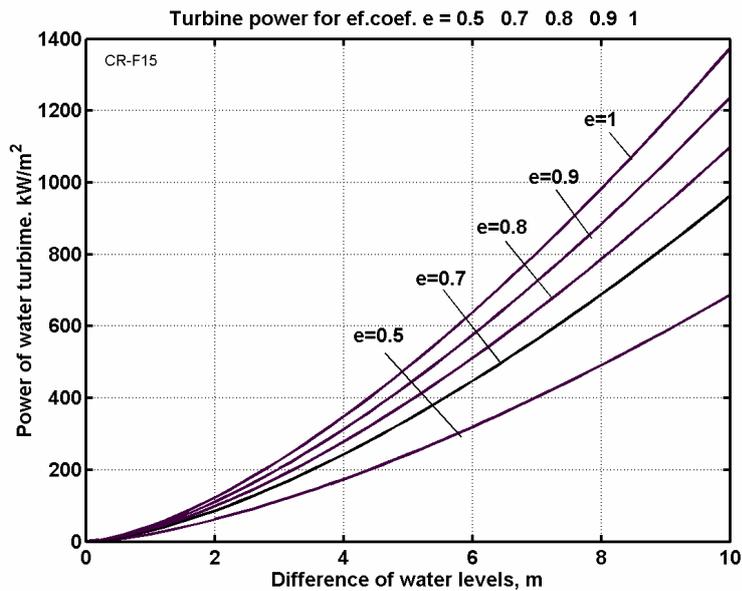

**Fig. 11**. Specific power of a water turbine via difference of water levels and turbine efficiency coefficient.

**3. Film thickness** is

$$\delta = \frac{g\gamma h^2}{2\sigma},\qquad(3)$$

where σ is safety film tensile stress, N/m². Results of computation are in Fig. 12. The fibrous material (Fiber B, PRD-49) has $\sigma$ = 312 kg/mm² and specific gravity $\gamma$ = 1.5 g/cm³.

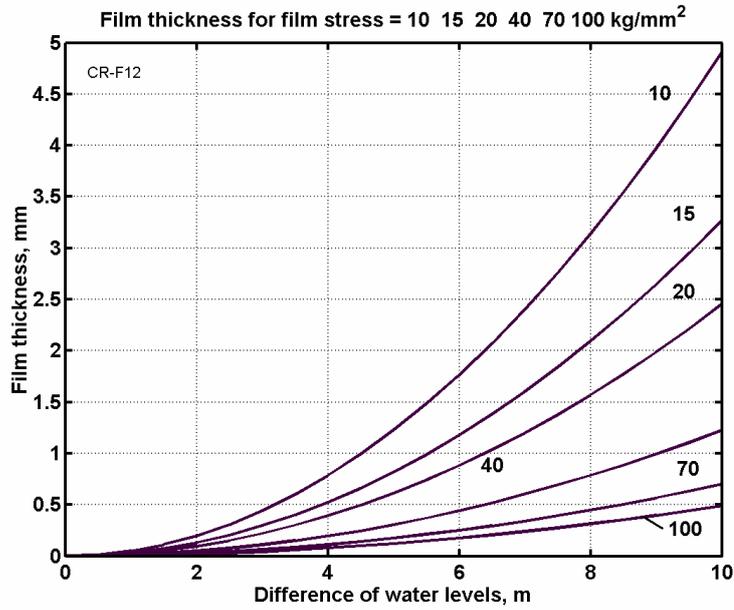

**Fig. 12**. Film (textile) thickness via difference of water levels safety film (textile) tensile stress.

4. **The film weight** of 1m width is

$$W_f = 1.2\delta\gamma H \qquad (4)$$

Computation are in Fig. 13. If our dam has long $L$ m, we must multiple this results by $L$.

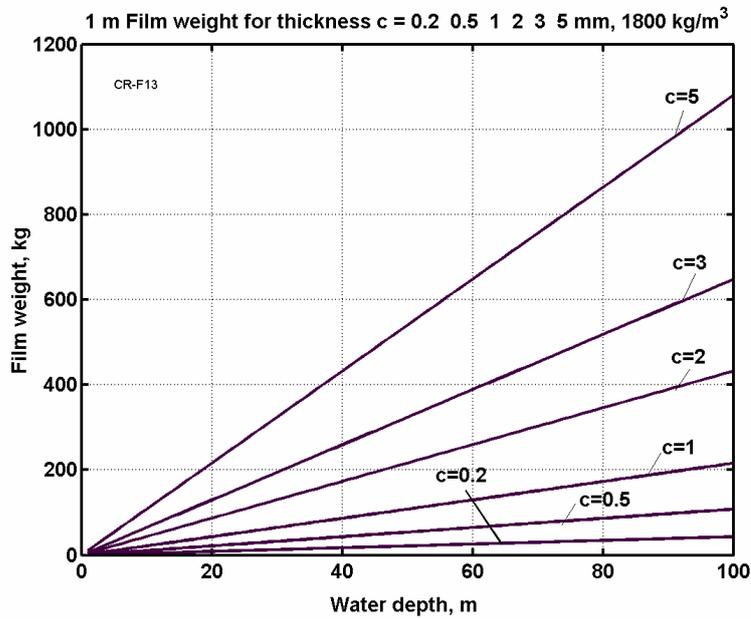

**Fig. 13**. One m Film weight via the deep of dam and film thickness $c$, density 1800 kg/m³.

5. **The diameter** $d$ of the support cable is

$$T = \frac{Pl_2}{2}, \quad S = \frac{T}{\sigma}, \quad d = \sqrt{\frac{4S}{\pi}}, \tag{5}$$

where $T$ is cable force, N; $l_2$ is distance between cable, m; $S$ is cross-section area, m$^2$. Computation is presented in fig. 14. The total weight of support cable is

$$W_c \approx 2\gamma_c HSL/l_2, \quad W_a = \gamma_c SL, \tag{6}$$

where $\gamma_c$ is cable density, kg/m$^3$; $L$ is length of dam, m; $W_a$ is additional (connection of banks) cable, m. The cheap current fiber has $\sigma = 620$ kg/mm$^2$ and specific gravity $\gamma = 1.8$ g/cm$^3$.

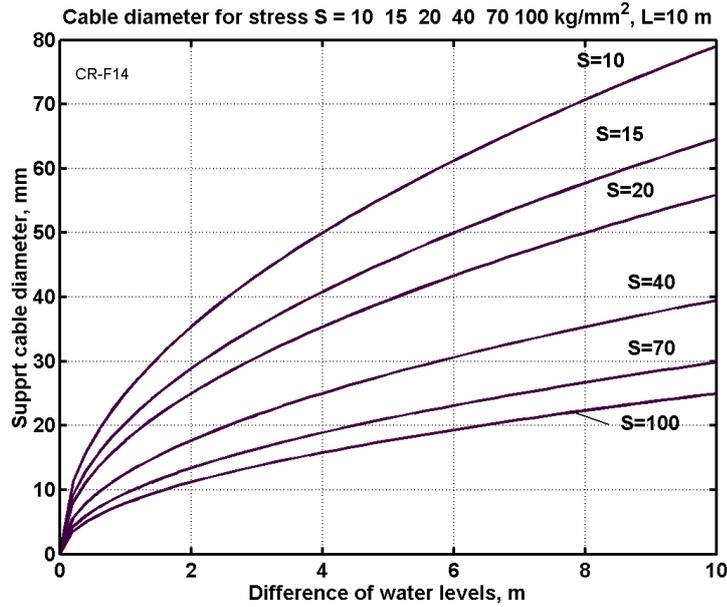

**Fig. 14**. Diameter of the support cable via difference of a water levels and the safety tensile stress for every 10 m textile dam.

**6. Maximum sea raise of water** from hurricane versus wind speed is

$$h = \frac{\rho V^2}{2\gamma}, \tag{7}$$

where $h$ is water raising, m; $\rho = 1.225$ kg/m$^3$; $V$ is wind speed, m/s; $\gamma = 1000$ kg/m$^3$ is water density.
Computation is presented in Fig. 15.

Wind speed is main magnitude which influences in the water raising. The direction of wind, rain, general atmospheric pressure, deep, and relief of sea bottom, Moon phase, also influence to the water raising and can decreases or increases the local sea level computed by Equation (7). For example, in hurricane "eye" the wind is absent, but atmospheric pressure is very low and sea level is high.

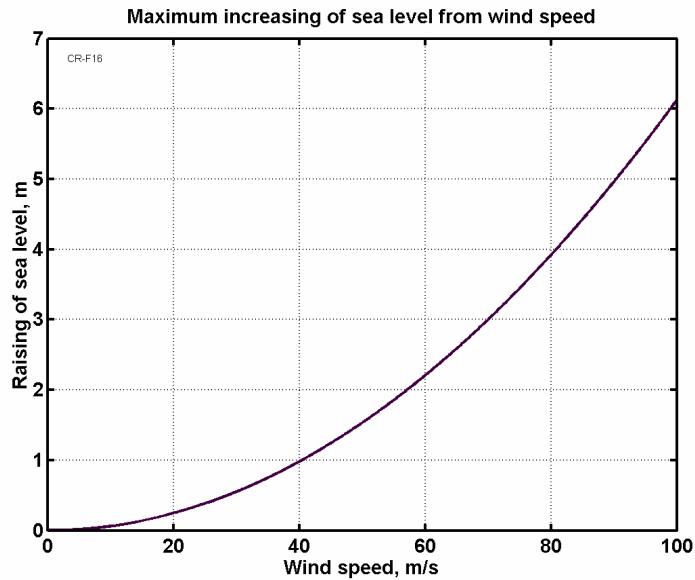

**Fig.15**. Raising of sea level via wind speed.

**Summary**

Authors offered and researched the new method and cheap design the land and sea textile (film) dams. The offered method of the protection of seaport cities against hurricane storm surge waves, tsunamis, and other weather related inundations is cheapest and has the very perspective applications for defense from natural weather disasters. That is also method for producing a big amount of renewable cheap energy, getting a new land for sea (and non-sea) countries. However, there are important details not considered in this research. It is recommended the consulting with author for application this protection.

**Application**

Using the graphs above, we can estimate the relevant physical parameters NYC dams and of many interesting macroprojects [25] - [30].

**Reflections**

The closest approximation of a TSSB at the Narrows is to be found in our arxiv.org postings "Ocean Terracing" (GOTO: physics/0701100) and "The Golden Gate Textile Barrier: Preserving California's Bay of San Francisco from a Rising North Pacific Ocean" (GOTO: physics/0702030). Here we examined similar proposed installation that seems obviously far more economical than what has been discussed in the media so far. Finally, we would like others to re-examine a Macro-engineering project proposed by Nigel Chattey—the FEASIBILITY STUDY OF THE ICCONN-ERIE INCORPORATED (1981).[24] Chattey suggested, among other macroprojects, the creation of a large area artificial island offshore of New York City. The presence of such an industrialized island, especially shaped to serve also as breakwater, may have very beneficial oceanographic effects on the Narrows!

**CITATIONS**